\documentclass{aa}

\newcommand{\prp}    {${\rlap.}^{\prime}$}
\newcommand{\grp}    {${\rlap.}^{\circ}$}
\newcommand{\pri}    {${\rlap.}^{\prime \prime}$}
\newcommand{\rl}     {${\rlap.}^{s}$}

\newcommand{\ltsima} {$\; \buildrel < \over \sim \;$}
\newcommand{\simlt}  {\lower.5ex\hbox{\ltsima}}            
\newcommand{\gtsima} {$\; \buildrel > \over \sim \;$}
\newcommand{\simgt}  {\lower.5ex\hbox{\gtsima}}            

\usepackage{graphicx}
\usepackage{txfonts}
\begin{document} 

\title{The star forming region Monoceros R2 as a gamma-ray source}
  

   \author{Mart\'{\i}, J.\inst{1}
   \and Luque-Escamilla, P. L.\inst{1}
   \and Mu\~noz-Arjonilla, A. J.\inst{1}
   \and S\'anchez-Ayaso, E. \inst{1}
   \and Munar-Adrover, P.\inst{2}
   \and S\'anchez-Sutil, J.~R.\inst{1}
   \and Romero, G. E.\inst{3,4}
   \and Paredes, J. M.\inst{2}
   \and Combi, J. A.\inst{3,4}
          }
   \institute{Escuela Polit\'ecnica Superior de Ja\'en, Universidad de Ja\'en, Campus Las Lagunillas, Edif. A3, 23071 Ja\'en, Spain\\
              \email{jmarti@ujaen.es, peter@ujaen.es, ajmunoz@ujaen.es, esayaso@ujaen.es, jrssutil@ujaen.es}
          \and
Departament d'Astronomia i Meteorologia, Institut de Ci\`encies del Cosmos (ICC), Universitat de Barcelona (IEEC-UB), 
Mart\'{\i} i Franqu\`es 1, E-08028, Barcelona, Spain\\
\email{pmunar@am.ub.es, jmparedes@ub.edu}
         \and
Instituto Argentino de Radioastronom\'{\i}a (IAR), CCT La Plata  (CONICET), C.C.5, (1894) Villa Elisa, Buenos Aires, Argentina \\
\email{romero@iar.unlp.edu.ar, jcombi@iar.unlp.edu.ar}
         \and
Facultad de Ciencias Astron\'omicas y Geof\'{\i}sicas, Universidad Nacional de La Plata, Paseo del Bosque s/n, 1900, La Plata, Argentina
             }
  

 
\abstract
   {After the release of the gamma-ray source catalog produced by the {{\it Fermi}} satellite during its first two years of operation, a significant
   fraction of sources still remain unassociated at lower energies. In addition to well-known high-energy emitters (pulsars, blazars, supernova remnants, etc.),
   theoretical expectations predict new classes of gamma-ray sources. In particular, gamma-ray emission could be associated
   with some of the early phases of stellar evolution, but this interesting possibility is still poorly understood.}
   {The aim of this paper is to assess the possibility of the {{\it Fermi}} gamma-ray source 2FGL~J0607.5$-$0618c being associated with the
   massive star forming region Monoceros R2.}
   {A multi-wavelength analysis of the Monoceros R2 region is carried out using archival data at radio, infrared, X-ray, and gamma-ray wavelengths.
   The resulting observational properties are used to estimate the physical parameters needed  to test the different physical scenarios.}
   {We confirm the 2FGL~J0607.5$-$0618c detection with improved confidence over
   the {{\it Fermi}} two-year catalog.  We find that a combined effect of the multiple young stellar objects
   in Monoceros R2 is a viable picture for the nature of the source.}
   {}

            
\titlerunning{The star forming region Monoceros R2 as a gamma-ray source}
\authorrunning{Mart\'{\i} et al.}

   \maketitle
%

\section{Introduction}

\object{Monoceros R2} (hereafter \object{Mon R2}) 
 is named after the se\-cond asso\-ciation of reflection nebulae in the constellation of Monoceros, i.e., the Unicorn,
following the nomenclature established in early studies \citep{origname}. More precisely, this designation usually refers
to a complex of active massive star formation embedded in a nearby ($\leq 1$ kpc), dense molecular core well below the Galactic plane ($l_{II}$=213\grp 7, $b_{II}$=$-$12\grp 6).
Strong thermal emission is probably associated with it in coincidence with the bright radio source \object{NVSS J060746$-$062303}  previously
detected in many other radio surveys.
A highly absorbed stellar cluster also exists here in coincidence
with NVSS J060746$-$062303. It lies close to the radio source's peak and next to the
center of a giant CO outflow \citep{bl83, w90}.
The distance to Mon R2 that we will use throughout this work is $\sim 830$ pc \citep{hr76}. This value is based on the distance modulus
resulting from fitting the zero age main sequence in color-color diagram of Mon R2 stars.

The Mon R2 central cluster content has been studied in detail in the near infrared by \citet{car97}
who suggest that it extends about 1.1 pc  $\times$ 2.1 pc  and contains $\geq$ 475 stars. The
estimated central density amounts to $\sim 9000$ stars pc$^{-3}$, with an average visual extinction $A_V \sim 33$ mag.
The most likely value for the ratio  of high mass (1 to 10 $M_{\odot}$) to low mass (0.1 to 1 $M_{\odot}$) stars
is about 0.11 and agrees well with expectations from a Miller-Scalo initial mass function. Most spectral classifications
of cluster members correspond to late types (G, K, and M). They often exhibit
infrared excesses in the  near infrared $K$ band suggestive of a significant  population of low-mass pre-main-sequence stars, i.e., mostly classical T Tauri stars. 
There is some indication
that the most massive cluster stars ($\sim 10$ $M_{\odot}$) should be located close to the center.
The earliest spectral type spectroscopically identified  corresponds to a B1 star \citep{car97} although, 
as discussed later in this paper, at least one O-type star
could be present as well based on the flux of ionizing photons required to account for the observed radio spectrum.
 An additional comprehensive review of the physical properties of the Mon R2 complex was published a few years ago  \citep{review}.
We refer the reader to this work and references therein for further details.

\begin{figure*}
   \centering
   \includegraphics[angle=0,width=18.0cm]{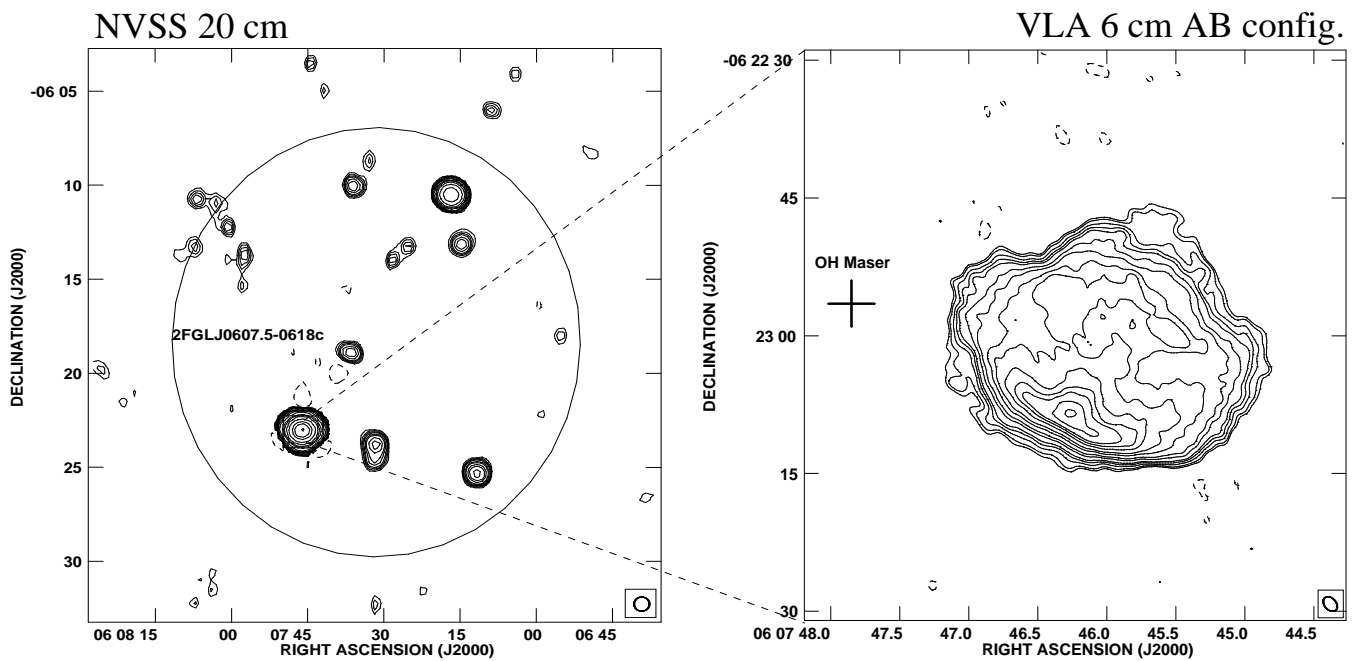}
      \caption{{\bf Left.} Radio map of the Mon R2 region from the NRAO VLA Sky Survey at the 20 cm wavelength.
      The restoring beam is a circular  $45^{\prime\prime}$ Gaussian. Contours shown
      correspond to $-3$, 3, 5, 6, 8, 10, 15, 20, 30, 50, 100, 200, 300, 500, 1000, 2000, 3000, and 5000 times 0.7 mJy beam$^{-1}$, the rms noise.
      The 95\% confidence ellipse for the gamma-ray source detected in the
      Mon R2 direction is plotted as provided by  the two-year 
      {\it Fermi} LAT catalog. Mon R2 is the brightest and most relevant radio source consistent with it.
      {\bf Right.} Zoom of the Mon R2 central region as observed with the VLA at the 6 cm wavelength. This high-resolution
      map has been selfcalibrated using the OH maser in the field (cross) and computed with pure uniform weight. Contours
      shown correspond to $-3$, 3, 4, 6, 8, 10, 15, 20, 30, 40, 60, 80, 100, 140, and 180
      times 0.7 mJy beam$^{-1}$, the rms noise. The restoring beam is shown at the panel bottom right corner
      as a 1\pri 80 $\times$  1\pri 25 ellipse, with position angle $41^{\circ}$. } \label{nvss_map}
   \end{figure*}

The source known as \object{IRS3} was discovered during historic infrared observations in the 1970s \citep{bw76} and proposed it to be a massive 
young stellar object (YSO).  It is the brightest infrared source in the Mon R2 field ($L_{{\rm 1-25 \mu m}} \sim 3 \times 10^3$ $L_{\odot}$)
 and currently considered a likely candidate to drive the giant CO outflow in the region. 
However, in modern  high angular resolution observations \citep{p2002} IRS3 does not appear to be a single object, 
but  resolved into several YSO components, three of them having estimated masses in the 5 to 15 $M_{\odot}$ range.
Based on their variable X-ray emission as detected by the {\it Chandra}
satellite, the same authors infer indirect evidence for magnetic interaction between two of these components and their respective surrounding disks.
The Mon R2-IRS3 system has also been reported to produce highly variable and flaring maser emission. In particular, spectacular maser flares of the hydroxyl (OH) molecule at the 
4765 MHz frequency have been recorded with changes in brightness of two orders of magnitude, and time scales of a few weeks to double in intensity \citep{oh98}.

Our interest in Mon R2 comes from theoretical models predicting gamma-ray emission associated
with YSOs under different physical scenarios. 
For instance, bipolar outflows from massive YSOs  can produce strong shocks when they interact with the surrounding medium
and can accelerate relativistic particles \citep{ara07,br2010}.  Similarly, 
acceleration of relativistic particles can occur as well in magnetic reconnection events in  T Tauri stars \citep{dv2011}. 
These new kinds of sources could broaden the domain of high-energy astrophysics beyond the study of traditional, well-known gamma-ray sources in the Galactic plane
such as pulsars, pulsar wind nebulae, supernova remnants, molecular clouds, gamma-ray binaries, etc.
In this context, a systematic cross-identification of gamma-ray sources 
from the two-year {\it Fermi} Large Area Telescope (LAT) catalog, hereafter 2FGL  \citep{2fgl},
 with known star forming regions yielded Mon R2 as a potential gamma-ray source candidate where some of these new physical scenarios
could be realized. The following sections in this paper are devoted to this purpose by assessing the observational evidence and theoretical expectations
in the Mon R2 case. The present work  complements other efforts to identify massive YSOs as potential counterparts of some galactic {\it Fermi} sources \citep{ma2011}.

\section{Observational data}

The unassociated {\it Fermi} LAT gamma-ray source \object{2FGL~J0607.5$-$0618c} lies in the direction of the Mon R2 complex and it was  included
as 1FGL~0608.1$-$0630c in the previous {\it Fermi} LAT catalog. In
the left panel of Fig. \ref{nvss_map}, this
coincidence is illustrated  by plotting the location of its  95\% confidence ellipse  onto 
a wide field radio image of the Mon R2 region retrieved 
from the  NRAO VLA Sky Survey, hereafter  NVSS  \citep{nvss}. 
Preliminary studies, based on the first {\it Fermi} catalog only,
raised doubts about this source being a possible spurious {\it Fermi} detection \citep{mah2010}. The same authors strongly supported a Galactic origin
in the alternative case of it being a true source. That the 2FGL catalog continues to include this detection, 
and without any classification flag warning, strongly suggests that this is the case.
Nevertheless, the observational parameters are still to some extent affected 
by the difficulties of accurate background modeling.  
Based on the 2FGL catalog,
the source currently appears detected at the  $7.6\sigma$ significance level. Its variability
index is 33.75, a value that does not reach the 41.64 threshold to be considered variable at the 99\% confidence level.
Significant fluxes are measured in the following energy bands: 
$F_{100-300~{\rm MeV}} = (1.4\pm0.4)\times 10^{-8}$ ph cm$^{-2}$s$^{-1}$,
$F_{300-1000~{\rm MeV}} = (6.75\pm1.4)\times 10^{-9}$ ph cm$^{-2}$s$^{-1}$, and
$F_{1-3~{\rm GeV}} = (1.2\pm0.3)\times 10^{-9}$ ph cm$^{-2}$s$^{-1}$. The corresponding spectrum
is well represented by a power law with spectral index $\gamma=2.39\pm 0.25$.

Is this  {\it Fermi} LAT source associated with Mon R2? To address this question we have undertaken a multi-wavelength approach to search
for possible counterparts at lower energies including radio, infrared, and X-ray. The different sets of data retrieved and calibrated for this
purpose, including an updated analysis of {\it Fermi} LAT data, are described in the following subsections. 

\subsection{Gamma rays}  \label{gamma_analisis}

   \begin{figure}
   \centering
   \includegraphics[angle=0,width=8.0cm]{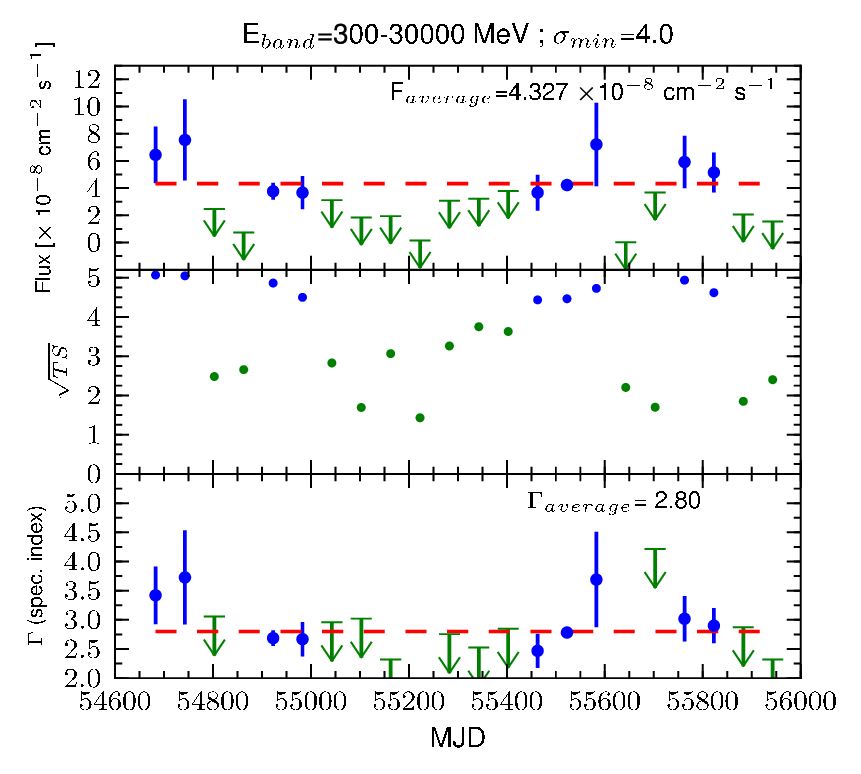}
      \caption{Light-curve study of the source 2FGL~J0607.5--0618c in the 0.3--300GeV energy range. 
      Upper pannel: \emph{Fermi} LAT light curve of 2FGL J0607.5--0618c sampled with 60-day bin intervals. 
      Middle pannel: value of $\sqrt{TS}$ for each light-curve bin. 
      Lower pannel: fitted spectral index $\Gamma$ for each light-curve bin. 
      Bins with $\sqrt{TS} <4.0$ ($\sigma_{min}$) show their flux value and spectral index as arrow upper limits. 
      Horizontal dashed lines represent average values.} \label{lightcurve}
   \end{figure}

  \begin{figure}
   \centering
   \includegraphics[angle=0,width=8.0cm]{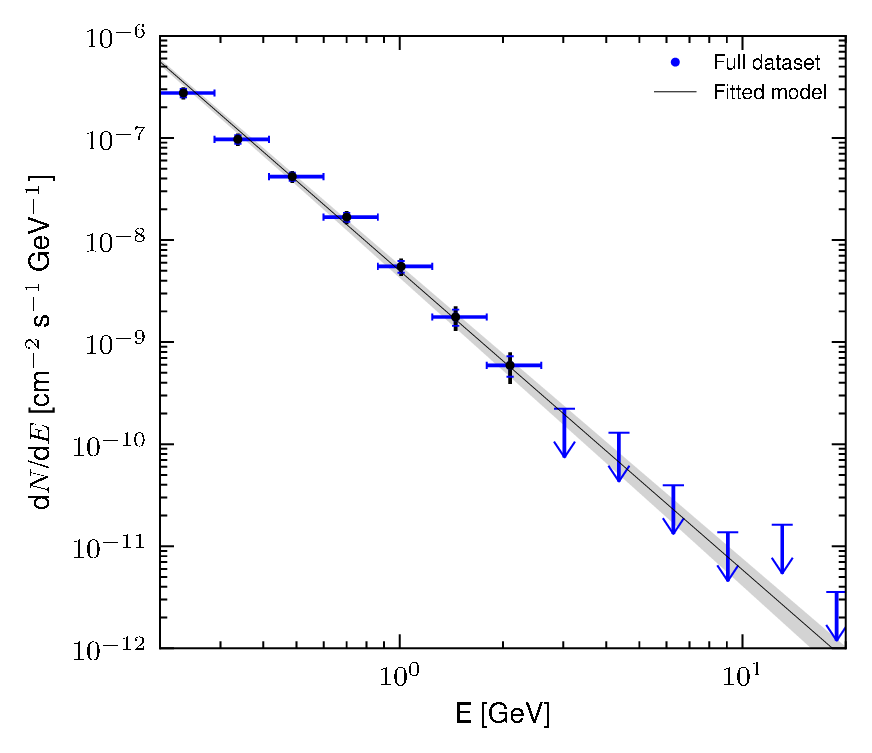}
      \caption{{\it Fermi} LAT spectrum of  2FGL~J0607.5$-$0618c resulting from our binned likelihood analysis.
      The shaded region represents the  final power-law fit and its uncertainty resulting from this work.
      Black error bars display the estimated systematic error.} \label{spectrum}
   \end{figure}

  \textit{Fermi} is a gamma-ray space telescope launched in June 2008. It carries two instruments, the Large Area Telescope (LAT) and the Gamma-ray Burst Monitor (GBM). 
  The LAT works in the 20 MeV to 300 GeV energy range and has a bigger collection area than any other past mission with more than 8000 cm$^2$. Its angular resolution is much better than previous gamma-ray telescopes, being less than 0\grp 15  for energies below 10 GeV. 
  The \textit{Fermi} LAT operates in scanning mode continuously observing the whole sky and thanks to its wide field of view  covers the full celestial sphere every few hours. 
We have reanalyzed all the \textit{Fermi} LAT data available at the time of writing around a 25$^{\circ}$  circle centered on 2FGL~J0607.5$-$0618c;
our main goals are to obtain an updated light-curve and the spectrum of our target source.
We are dealing here with 3 years and 8 months of data from 2008 August 4th to 2012 April 13th.
This almost doubled the observation time of the present 2FGL catalog. 
For this purpose, we used the Science Tools provided by the \textit{Fermi} satellite team. The version of the Science Tools used was {\tt v9r27p1} with the P7SOURCE\_V6
instrument response function (IRF). The reader is referred to  \textit{Fermi} instrumental publications  for further details about IRFs and other  calibration details \citep{irfs}.
We have adopted the current Galactic diffuse emission model ({\tt gal\_2yearp7v6\_v0.fits}) in a likelihood analysis and {\tt iso\_p7v6source.txt} as the isotropic model;
 the point source catalog {\tt gll\_psc\_v07.fit} has been used\footnote{{\tt http://fermi.gsfc.nasa.gov/ssc/data/access/}}. In the modeling of the data, the Galactic background and diffuse components remained fixed. 
 We selected  Pass7 Diffuse class events with energies between 0.2 and 300 GeV. Among them, we limited the reconstructed zenith angle to be less than 105$^{\circ}$ to greatly reduce gamma rays coming from the limb of the Earth's atmosphere. We selected the good time intervals of the observations by excluding events that were taken while the instrument rocking angle was larger than 52$^{\circ}$. 
 
 To get both the light-curve and spectrum, we used a power-law model based on the 2FGL catalog
  for all sources within a 25$^{\circ}$ radius around the position of our target 2FGL~J0607.5$-$0618c.
 Centered on it, we also defined a closer region of interest (ROI) covering only 10$^{\circ}$. The data reduction procedure is always based on a likelihood analysis in two steps.
 The first step used the Minuit optimizer to fit both the amplitudes and spectral indices of sources within the ROI, and only the amplitudes for those outside it.
 The second step involved the Newminuit optimizer for fitting only the spectral indices for those sources inside the ROI, while the rest of parameters remained fixed.
 At all times, the central target source 2FGL~J0607.5$-$0618c kept all its parameters free.
  
To obtain the light-curve we divided our data sample into 60-day bins and performed a full unbinned likelihood analysis on each of them. 
On the other hand, to get the source spectrum we binned our sample by energy and carried out a binned likelihood analysis for each energy bin.
The final results of these analyses can be seen in Figs.  \ref{lightcurve} and \ref{spectrum} for the light curve and spectrum, respectively.
 In Fig. \ref{lightcurve}, errors in flux and in the spectral index $\Gamma$ are calculated through the covariance matrix of all the free parameters in the model. 
 The likelihood analysis gives also a test statistic (TS) value, whose square root indicates the significance of the signal that we get from the position of the source. 
 There are only 9 bins out of 21 with a $\sqrt{TS} > 4.0$.  
In the Fig. \ref{spectrum} spectrum, the source is not detected at energies above $\sim$3 GeV.
The final fitted model is also plotted.
It has a spectral index $\gamma = -2.73 \pm 0.09$, steeper than the one in the 2FGL catalog.
This yields an energy flux of $(2.20 \pm 0.25) \times 10^{-8}$  ph cm$^{-2}$ s$^{-1}$, (TS$=$154.25, equivalent to 12.42$\sigma$)
in the 0.2-300 GeV band.
At the Mon R2 distance, this is equivalent to a gamma-ray luminosity of $1.2 \times 10^{33}$ erg s$^{-1}$.

All errors quoted up to this stage are statistical only. Systematic errors are a delicate issue to assess and their calculation may be complicated. 
The {\it Fermi} collaboration uses bracketing IRFs, 
which vary in effective area and from them they calculate the systematic effect on the analysis results. Unfortunately, these IRFs are not public and we cannot estimate the systematic errors in this way.
Typically,  the systematic uncertainty is found to follow more or less the statistical uncertainty and is of the same order \citep{2fgl}, i.e., it is larger for fainter sources in relative terms.
 More precisely, the dispersions of flux and spectral index are $0.8\sigma$ for sources with Galactic latitude $\left|b\right| < 10^{\circ}$ 
 (2FGL~0607.5$-$0618c is at Galactic latitude $b=-$12\grp 6).
The systematic error in the flux calculation depends on $x=\log(E/{\rm MeV})$ and
amounts to 10\% for $x=2$, 5\% for $x=2.75$ and 20\% for $x=4$ \citep{abdo09}. 
We have linearly interpolated these values to account for the systematic uncertainty  in the spectral points shown in Fig. \ref{spectrum} (black error bars).

\begin{figure}
   \centering
   \includegraphics[angle=0,width=8.0cm]{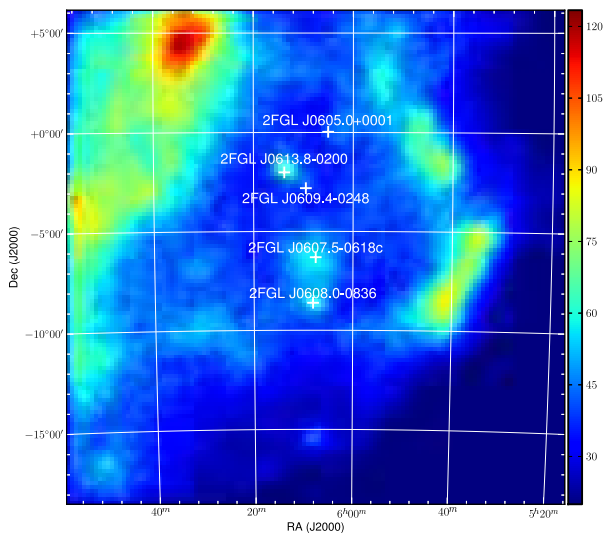}
      \caption{Image of the  2FGL~J0607.5$-$0618c field resulting from the {\it Fermi} LAT data processed in this work. 
     The image has been smoothed with a 2-pixel Gaussian kernel.  
      The target centroid position is marked at the center as a green cross.
      Other sources from the 2FGL catalog are also marked for image completeness.}
    \label{count_map}
   \end{figure}

In addition, we produced a gamma-ray image of the region containing the source 2FGL~J0607.5$-$0618c (see Fig. \ref{count_map}).
The centroid of its counts excess has been estimated using the {\tt gtfindsrc} function. The source best-fit J2000.0 position 
 is now  right ascension $06^h08^m10.87^s$ and declination $-06^{\circ}29^{\prime}32.8^{\prime\prime}$
 with an error circle of $0.19^{\circ}$, which is displaced $0.25^{\circ}$ from the original 2FGL position.  The Mon R2 central region is still well inside our new error circle.
To study possible deviation from the point-like source we used the {\tt gttsmap} function to generate two TS-maps of the ROI,
one of them with a model containing 2FGL~J0607.5$-$0618c and the other without it. 
The residuals after subtracting these two TS-maps show that our source is not extended. 
We also checked for extension by producing a model map using the {\tt gtlike}  task with the output model from the binned likelihood analysis, 
and subtracting it from a counts map generated with {\tt gtbin}. Again, the residuals map indicates no signal of extension for 2FGL~J0607.5$-$0618c.

\subsection{Radio}

%
\begin{table}
\caption{NVSS continuum radio sources inside the 95\% ellipse of  2FGL~J0607.5$-$0618c}             
\label{nvss_sources}      
\centering                          
\begin{tabular}{c c c}        
\hline\hline                 
NVSS  & 20 cm flux density & Remarks(*) \\    
 name   & (mJy)      &     \\
\hline                        
\object{J060655$-$061803}  &   $4.0 \pm 0.6$ &             \\
\object{J060658$-$062209}  &  $2.6 \pm 0.5$ &             \\
\object{J060711$-$062519}  & $28.6 \pm 1.3$  &             \\
\object{J060714$-$061308}  & $11.5 \pm 1.1$  &             \\
\object{J060716$-$061030}  & $143.7 \pm  5.1$ &  Non-thermal           \\
\object{J060725$-$061313}  &  $ 4.8 \pm 0.5$ &             \\
\object{J060728$-$061359 } & $4.8 \pm 0.5$ &    Optical counterpart         \\
\object{J060731$-$062356 } &  $27.4 \pm 1.5$ &   BD $-$6 1415?          \\
\object{J060732$-$060843}  &   $4.1 \pm 0.5$  &             \\
\object{J060735$-$061002}  &  $10.0\pm 0.5$  &             \\
\object{J060736$-$061852}  &  $11.4 \pm 1.0$  &             \\
\object{J060746$-$062303}  & $4110.2 \pm 144.1$  &   Mon R2 cluster          \\
\object{J060757$-$061347}  &    $5.9 \pm 0.6$ &   X-ray/IR counterpart  \\
\object{J060757$-$061522}  &   $3.0 \pm 0.5$  &             \\
\object{J060759$-$062152}  &    $2.2 \pm 0.5$ &             \\
\object{J060759$-$061212}  &  $11.4 \pm 1.8$ &             \\
\object{J060802$-$061057}  &   $4.8 \pm 0.6$ &             \\
\object{J060807$-$061316}  &   $3.9 \pm 0.5$ &             \\
\object{J060810$-$061353}  &   $3.7\pm 0.7$  &             \\
\hline                                   
\end{tabular}
~\\(*) See discussion.
\end{table}

The list of NVSS radio sources displayed
in the left panel of Fig. \ref{nvss_map} consistent with the {\it Fermi} LAT source position is given in Table \ref{nvss_sources}.
In addition, we also searched the NRAO archives for higher angular resolution radio observations. 
Only one project (AE0167, observer Simon Ellingsen) in the hybrid AB configuration appeared suitable for our goals with a total on-source time of 2.6 h. 
The observation was carried out on 2007 October 24th at the 4765.4 MHz frequency using a spectral model with 512 channel covering a 0.8 MHz bandwidth.
 Only right circular polarization was recorded. \object{3C 286} was used as amplitude and bandpass calibrator, while the nearby \object{0607$-$085}
  was observed as phase calibrator.
 The bootstrapped flux density of the phase calibrator was estimated to be $1.26\pm0.01$ Jy.

 The OH maser is well detected in this VLA experiment as a compact source with a very bright flux density of 34 Jy. 
Its measured J2000.0 position   turns out to be right ascension 06$^h$07$^m$47\rl 850$\pm$0\rl 001 
and declination $-06^{\circ} 22^{\prime}$ 56\pri 50$\pm$0\pri 01,
 in agreement with previous measurements with the MERLIN interferometer \citep{oh98}.
 This detection enabled us to self-calibrate the maser channels
 and transfer the calibration solution to the continuum visibility data to produce an excellent high-fidelity map. 
 The cross-calibration procedure used here is the same  that we used successfully  in previous YSO radio studies  \citep{croscalib}.
 The final result is presented as a contour plot
 in the right panel of Fig. \ref{nvss_map}, where the cross symbol marks the IRS3 OH maser position. Only extended emission is detected and no compact
 components emitting in the continuum are brighter than 2.1 mJy (3$\sigma$).
 This VLA map is also used as the red layer of the Mon R2 tri-chromatic
 image displayed in Fig. \ref{tricromia}.
 
%
\begin{table}
\caption{Radio flux densities of Mon R2}             
\label{radiofd}      
\centering                          
\begin{tabular}{c c c c}        
\hline\hline                 
Frequency & Flux Density & Reference  \\    
(GHz)         &    (mJy)         &                   \\
\hline                        
 0.365  &   $602 \pm  120$  & SPECFIND V2.0 \\
 0.408  &   $760  \pm 150$  & SPECFIND V2.0\\
 1.400  &  $4110 \pm  820$ & SPECFIND V2.0\\
 2.700  &  $5600 \pm 1100$ & SPECFIND V2.0\\
            &  $9800 \pm 2000$ & SPECFIND V2.0\\
 4.850 & $ 6730 \pm 1300$ & SPECFIND V2.0 \\
 8.440 & $7075  \pm   140$ & VLA AP0238(*) \\
\hline                                   
\end{tabular}
~\\(*) 2\% absolute flux calibration error assumed.
\end{table}

 \begin{figure}
   \centering
   \includegraphics[angle=0,width=9.0cm]{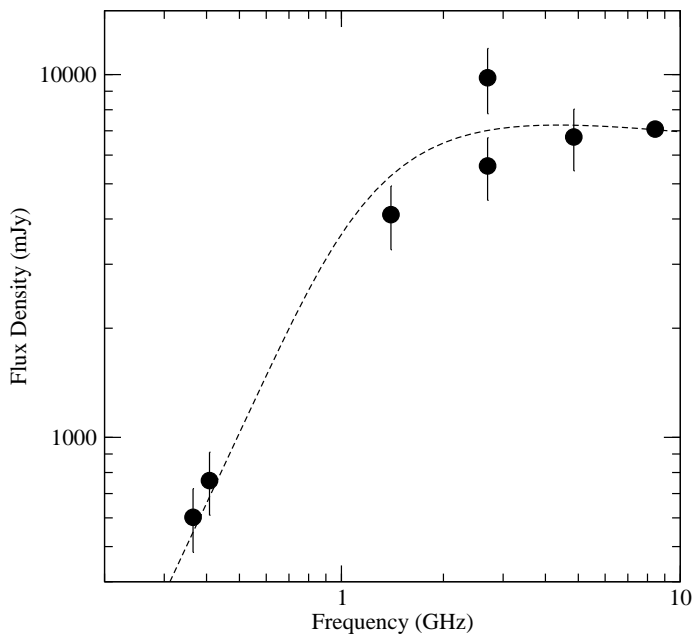}
      \caption{Radio spectrum of the Mon R2 central region, where the absorbed cluster is located, assembled by collecting detections from different surveys
      included in the SPECFIND V2.0 catalog supplemented with an archival 8.4 GHz detection with the VLA.
      The dashed line is a free-free thermal spectrum fit  from Eq. \ref{ff_fit}.} \label{radio_spectrum}
   \end{figure}
     
Early measurements \citep{dw75} indicated that radio emission from the Mon R2 central region was of
 thermal, free-free nature. To confirm this, in Fig. \ref{radio_spectrum} we assembled detections at different frequencies retrieved from the
SPECFIND V2.0 catalog \citep{specfind}. The plot also includes a 8.4 GHz detection obtained in 1992 from
an archival VLA observation in its most compact D configuration (Project Id. AP0238, observer Patrick E. Palmer).
The corresponding visibility data was also downloaded from the NRAO archive and calibrated by us following standard
procedures for simple continuum interferometric data.
All the individual flux densities are given numerically in Table \ref{radiofd}. A least-squares fit of a theoretical thermal free-free
model \citep{vk88} provides a good representation of the data (see Fig. \ref{radio_spectrum}), with the following flux density dependence as a function of frequency
\begin{equation}
\left[ \frac{S_{\nu}}{{\rm mJy}} \right] = (4113 \pm 438) \left[\frac{\nu}{{\rm GHz}}\right]^2 
\left(1 - e^{-(2.15\pm 0.24) \left[\frac{\nu}{{\rm GHz}}\right]^{-2.1}}\right).  \label{ff_fit} 
\end{equation}
This expression will be used later for discussion purposes. 
We also remark here that weak non-thermal sources, at tens
of mJy level, cannot be detected in our data against the strong thermal dominant emission.

\subsection{Infrared}

 \begin{figure*}
   \centering
   \includegraphics[angle=0,width=18.0cm]{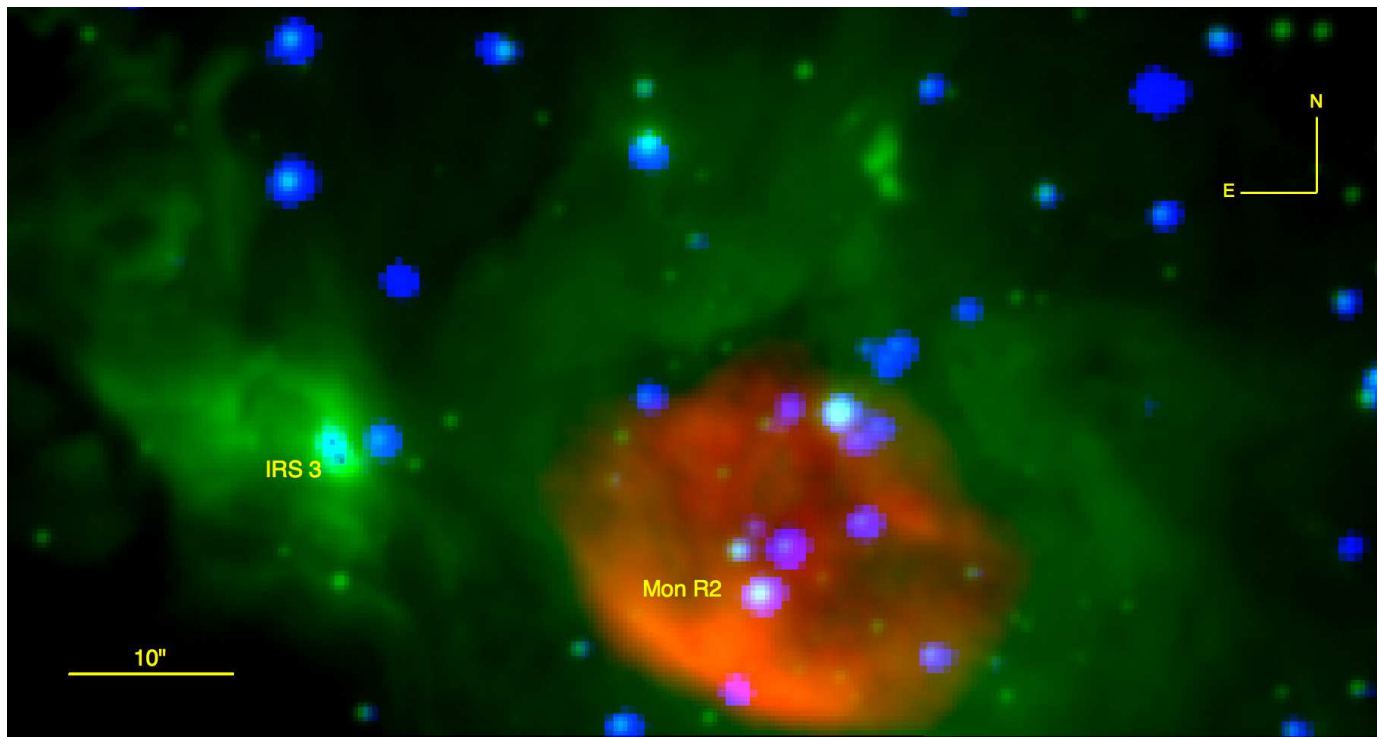}
      \caption{Trichromatic image of the Mon R2 central cluster region. Red, green, and blue colors correspond to
      VLA 6 cm radio, ESO NTT $Ks$-band infrared, and {\it Chandra} X-ray emission, respectively. The brightest infrared
      and maser source in the field (IRS3) is also indicated. The field of view shown is
       fully inside the 95\% confidence ellipse for 2FGL~J0607.5$-$0618c. It  corresponds to
       $0.75 \times 1.40$ arcmin$^2$ with north up
      and east left.} \label{tricromia}
   \end{figure*}

We reanalyzed ESO archival images at near-infrared wavelengths of the 
Mon~R2 field in the $J$, $H$, and $K_s$ bands obtained with the 
ESO New Technology Telescope (NTT) and the SOFI instrument on 
2001 March 6th  (Project Id. 66.C-0310). Standard procedures were followed for sky 
background subtraction, flat-fielding, and median combining of 
individual frames using the IRAF software package. Astrometry was 
calculated based on the data of 35--40 stars within the field 
whose accurate positions were retrieved from the 2MASS catalog. The 
total combined astrometric error is below 0\pri 1.
Finally, the $K_s$ band image was selected as the green layer of the tri-chromatic image displayed in Fig. \ref{tricromia}
where the object labelled as IRS 3 stands outs as the most luminous at infrared
wavelengths ($L_{{\rm 1-25 \mu m}} \sim 3 \times 10^3$ $L_{\odot}$).

\subsection{X-rays}

The {\it Chandra} Satellite carried out one observation of Mon R2 on 2007 June 19th ( ObsID 1882 ) at
a nominal J2000.0 pointing direction of $\alpha$=06$^h$07$^m$50\rl 63, $\delta=-06^{\circ}22^\prime$50\pri 45
by the Advanced CCD Image Spectrometer (ACIS) camera with a $17^{\prime} \times 17^{\prime}$ field of view. This  observation, fully covering
the central cluster region but not the whole 95\% confidence {\it Fermi} LAT ellipse, was re-calibrated using the CIAO (version 4.3) and CALDB (version 4.4.3) software packages.  
The light curves of photons above 10 keV were extracted from the entire field of view of the camera  to exclude strong background 
flares and we also exclude intervals up to 3 $\sigma$ to produce a good time interval (GTI) file.
The resulting X-ray image was smoothed with a kernel of 2 pixels and used as
the blue layer of the tri-chromatic image displayed in Fig. \ref{tricromia}.

Our subsequent re-analysis is essentially consistent with those already published  \citep{Chandra_japo, Chandra_pasj},
where satisfactory spectral fits are often achieved using thermal plasma models common for YSOs. No X-ray sources with
clear evidence for a non-thermal (e.g., power-law) spectrum have been detected in Mon R2.

\section{Discussion}

From the analysis of all {\it Fermi} LAT data available in Sect. \ref{gamma_analisis}, the gamma-ray source
2FGL~J0607.5$-$0618c has been confirmed with higher confidence (about $12\sigma$) than the not-so-significant
detection in the 2FGL  catalog. While the spectral properties remain comparable to those anticipated by the 2FGL catalog,
 the light curve resulting from the whole {\it Fermi} database analysis 
 is consistent with a steady level of emission.
Based on the reality of 2FGLJ0607.5$-$0618c as a starting point,
in this section we explore different physical scenarios where the high-energy emission
could originate in Mon R2, i.e., the outstanding and nearby massive star forming region in the direction of this gamma-ray source.
The possibility of a different counterpart is also assessed in Sect. \ref{end}.

\subsection{Mon R2 as a passive target for cosmic rays}

In this case, the gamma-ray emission is mainly dependent on the Mon R2 density.
The radio spectrum in Fig. \ref{radio_spectrum} enables us to estimate both the density
and mass of ionized gas which is a lower limit to the total value. For this purpose,
we first consider the detailed dependence of the radio flux density as a function
of frequency for thermal free-free radio emission
\begin{equation}
S_{\nu} = \frac{2 k \nu^2}{c^2} \Omega_s T_e (1-e^{-\tau_{ff}}),
\end{equation}
where the optical depth can be expressed as
\begin{equation}
\tau_{ff}=0.082 
                          \left[\frac{EM}{{\rm cm^{-6} pc}}\right]
                        \left[\frac{T_e}{{\rm K}}\right]^{-1.35}
                     \left[\frac{\nu}{{\rm GHz}}\right]^{-2.1},
\end{equation}
$k$ being the Boltzmann constant and $c$ the vacuum speed of light.
                         
 Approximating the solid angle $\Omega_s$ of the 
radio emitting region by a spherical source with angular diameter 0\prp 5, the numerical coefficients in Eq. \ref{ff_fit}
imply the following estimates of the
electron temperature $T_e \simeq 9700$ K, emission measure $EM\simeq 6.3 \times 10^6$ cm$^{-6}$ pc,
electron density $n_e \simeq 7.5 \times 10^3$ cm$^{-3}$, and a mass of ionized gas amounting to
1.0 $M_{\odot}$. This value represents a small fraction of the total mass in the central $4^{\prime} \times 4^{\prime}$ Mon R2 core estimated to be
about $1000$ $M_{\odot}$ by radio observations of molecular lines \citep{t1997}.

It is also possible to derive the flux of ionizing photons needed to keep the radio nebula ionized
that turns out to be $2.8 \times 10^{48}$ ph s$^{-1}$. 
This value could be sustained by a single late O-type star. Although its existence in Mon R2 is very likely,  
the earliest spectral type identified so far in Mon R2, as stated in the Introduction, corresponds to a B1 star (see also Table \ref{spectral_type}).

The ambient gas in Monoceros can be irradiated by cosmic rays and produce $\gamma$-ray emission by neutral pion decays generated in proton-ion interactions. 
The expected total flux is \citep{aha1991, com1995,tor2003}

\begin{equation}
F_\gamma= \frac{1}{4\pi d^2} \int_{V_0} n(\bar r) q_\gamma ({\bar r}) d^3{\bar r},
\end{equation}
where $q_{\gamma}$ is the $\gamma$-ray emissivity, $d$ the distance to the cloud, and $V_0$ is volume. 
Neglecting all gradients within the cloud, this equation reduces to
\begin{equation}
F_\gamma= \frac{M_{{\rm cl}}}{m_p} \frac{q_\gamma}{4\pi d^2},
\end{equation}
where $M_{{\rm cl}}$ is the mass of the cloud. We can re-write this as
\begin{equation}
F_\gamma  \sim 10^{-9} \left(\frac{M_{\rm cl}}{10^3 M_{\odot}}  \right) \left( \frac {d}{{\rm kpc}}\right)^{-2} 
\left(\frac{q_\gamma}{q_{_{-25}}}  \right) \; {\rm ph\; cm^{-2} s^{-1}} \label{hadro}
\end{equation}
with the constant $q_{_{-25}}$ being $10^{-25}$ s$^{-1}$ (H -- atom)$^{-1}$. 
The factor $q_\gamma$ will be enhanced in comparison with its normal value in case of a local source of relativistic particles. In a
passive giant molecular cloud exposed to the same proton flux measured at the Earth,
the $\gamma$-ray emissivity above 100 MeV is 
$q_{\gamma,0}=1.53  \ \eta \  q_{_{-25}}(\geq 100 \, \rm MeV) \rm (H - atom)^{-1}s^{-1}$, where the parameter $\eta  \simeq 1.5$ takes into account
the contribution of nuclei both in cosmic rays and in the
interstellar medium \citep{der1986,aha2001}. If the
shape of the cosmic ray spectrum in the cloud does not differ much from
that existing in the Earth's neighborhood, we can approximate
\begin{equation}
\frac {q_\gamma}{q_{\gamma,0}} \sim \frac{\epsilon_{\rm
CR}}{\epsilon_{\rm CR,0}} \sim k_{\rm s} ,
\end{equation}
where $\epsilon_{\rm CR}$ and $\epsilon_{\rm CR,0}$ are the cosmic ray energy densities in the cloud and the solar vicinity respectively, while
$k_{\rm s}$ is a cosmic ray enhacement factor.

Using the values of $M_{\rm cl}$ and $d$ for Mon R2, and the observed $\gamma$-ray ($E \geq 100$ MeV) flux of 
$\sim 2.2\times 10^{-8}$ ${\rm ph\; cm^{-2} s^{-1}}$ in Eq. \ref{hadro}, we find that at least a cosmic ray enhancement of  $k_{\rm s} \sim 7$
must exist in the region.
 Given such a large value, we infer that the
 gamma-ray emission observed is not likely to be the result of the Mon R2 region acting as a passive target for the Galactic sea of cosmic rays.

%
\begin{table*}
\caption{Stars spectroscopically classified in the Mon R2 cluster($^a$)}             
\label{spectral_type}      
\centering                          
\begin{tabular}{cccccccccccccc}        
\hline\hline                 
Spectral type &  O5-9  & B0-5 & B5-9 & A0-5 & A5-9  & F0-5 & F5-9 & G0-5 & G5-9 & K0-5 & K5-9 & M0-5 & M5-9 \\    
\hline
No. of stars          &     1($^b$) &    1      &    1     &     1      &  $-$     &    1     &    $-$    &    2      &     3      &    7     &    5      &      7      &     2     \\
\hline                        
\end{tabular}
~\\(a) Compilation based on Table 3 in Carpenter et al. (1997).\\
(b) Inferred from radio continuum spectrum.\\
\end{table*}

\subsection{IRS3 as massive YSOs with gamma-ray emission}

It is known that the infrared source IRS3 in Mon R2 splits into at least six components (named A to F) when observed with high angular
resolution,  using techniques such as speckle imaging \citep{p2002}, for example. In Fig. \ref{tricromia}, the brightest A and B components are  well resolved in the $Ks$-band layer.
Based on our infrared astrometry, component A is  also very  consistent with the strong OH maser spot in the field to the $\pm$0\pri 1 level.
In high angular resolution images \citep{p2002}, 
A and B appear surrounded by elongated extended emission in the infrared at sub-arcsecond angular scales.
Preibisch et al. (2002) also estimate their masses  to be above $\sim 10$ $M_{\odot}$. All these facts suggest that the two brightest IRS3 components are probably massive YSOs powering
collimated outflows.  Moreover, the outflow in the B case is particularly visible in their images with a knotty structure \citep{p2002}. 

As mentioned in the introduction, massive YSOs with outflows produce shocks and accelerate cosmic rays. Recently, polarized radio emission was detected in the outflows of HH80-81, a well-known massive YSO \citep{cg2010}.
The total output in cosmic rays of a massive YSO deeply embedded in a molecular cloud is estimated to be $\sim 10^{35}$ erg s$^{-1}$
\citep{ara07,rom08,br2010}.  
These cosmic rays can produce around $10^{33}$ erg s$^{-1}$ in gamma rays through $pp$ interactions with the ambient gas. The contribution from relativistic electrons will depend on the dominant losses for these particles. In a dense medium, relativistic Bremmstrahlung dominates yielding a similar figure to the $pp$ interactions (because of the similarity of the cross sections).

The relevant loss timescales for electrons, $t_{\rm loss}=-E/\dot{E}_{\rm loss}$, 
are given by the  expressions (Bosch-Ramon et al. 2010)
\begin{equation}
t_{\rm synchr}\approx 4\times 10^{11}\,B_{-3}^{-2}\,E_{\rm GeV}^{-1}\,{\rm s}\,,
\end{equation}
\begin{equation}
t_{\rm IC}\approx 1.6\times 10^{13}\,u_{\rm IR-9}^{-1}\,E_{\rm GeV}^{-1}\,{\rm s}\,,
\end{equation}
\begin{equation}
t_{\rm Bremsstr}\approx 3.5\times 10^{11}\,n_{\rm 3}^{-1}\,\,{\rm s}\,,
\end{equation}
where $B_{-3}$ is the magnetic field strength in mG, $u_{\rm IR-9}=u_{\rm IR}/10^{-9}\,{\rm erg~cm}^{-3}$ is the 
energy density of the IR photon field and $n_{3}$ the particle density in units of $10^3$ cm$^{-3}$. We see then that IC losses will be negligible when compared to Bremsstrahlung at $E\sim 1$ GeV in a standard cloud (the normalizations are chosen to reflect average values).  Altogether, the predicted emission from pp and relativistic Bremsstrahlung is of the order of the
inferred gamma-ray luminosity of $1.2 \times 10^{33}$ erg s$^{-1}$.




\subsection{T Tauri stars in Mon R2 as possible gamma-ray sources}

A significant number of T Tauri stars are present in the Mon R2 region.  
Nearly one hundred YSOs of this kind have been classified among sources with X-ray and infrared counterparts including
classical (25\%) and weak-lined (75\%)  T Tauri stars \citep{Chandra_pasj}.

The magnetosphere of T Tauri stars is expected to produce gamma-ray emission \citep{dv2011} at a moderate level of $\sim 10^{31-32}$ erg s$^{-1}$. 
This radiation is the result of particle acceleration up to relativistic energies by fast and turbulent magnetic reconnection events in the magnetosphere 
of the protostars. A condition for these events to occur is a relatively high magnetic field, so the Alfven speed can be high as well, 
allowing  a maximum energy beyond 100 GeV for protons (del Valle et al. 2011, and references therein). 
The X-ray activity of T Tauri stars  suggests the existence of surface magnetic fields of $\sim 10^2$~G \citep{fm1999}.
Although these fields cool down the electrons through synchrotron radiation, protons are ejected and interact with the matter in the environment \citep{dv2011}.

If each protostar produces about $5 \times 10^{31}$ erg s$^{-1}$, a hundred of them packed in a small region could radiate as much as
$\sim 5 \times 10^{33}$ erg s$^{-1}$ at energies above 200 MeV. We conclude that the accumulative effect of the large number
of the protostars present in the Mon R2 region might represent a significant contribution to the overall emission detected
by {\it Fermi}.

\subsection{Runaway stars}

The arc-shaped bowshocks created in the interstellar medium by runaway massive stars have been proposed as acceleration sites
of relativistic particles and hence as a possible new kind of gamma-ray source \citep{dvr12}. We inspected GLIMPSE infrared images
of the Mon R2 region searching for this kind of structure but found none. The structures, however, are difficult to find. 
A recent survey \citep{peri2011} detected only 28 bowshocks out of almost 300 stars inspected. We cannot rule out at present
some level of relativistic particle injection in Mon R2 by bowshocks. Although electrons are expected to cool close to the stars by 
inverse Compton losses, protons tend to be convected away and then can mix with cloud material \citep{ben2010, dvr12},
hence contributing to 2FGL~J0607.5$-$0618c. In addition, several bowshocks formed by the jets of low-mass protostars are detectable in radio. Although their efficiency to accelerate cosmic rays should be modest, their global contribution might help to enhance the relativistic particle population inside Mon R2. 
We note here that a few tens of thermal jets from low-mass stars  
are actually visible in $H_2$ infrared images \citep{h2_jets} inside the {\it Fermi} LAT ellipse.

\subsection{A counterpart unrelated to Mon R2?}  \label{end}

The possibility of an unrelated  coincidence of  2FGL~J0607.5$-$0618c with Mon R2 occurring needs to be taken into account to ensure
that the physical scenarios discussed above are indeed conceivable.
To estimate this possibility we followed a Monte Carlo formalism \citep{mc} originally developed
for EGRET gamma-ray sources. In short, we simulated a large number  of {\it Fermi} LAT synthetic populations with the same
properties of the two-year year catalog and searched for random coincidences with Mon R2 at $1^{\circ}$ and $2^{\circ}$ binning.
For runs with more than $10^4$ populations (we ran up to $10^6$), a probability of chance association of
$\sim 10^{-3}$ is consistently found.

Despite the low value found above, we also searched the {\it Fermi} LAT 95\% confidence ellipse for other objects that could provide
alternative counterpart candidates to the gamma-ray emission without a direct connection with the Mon R2 star forming region. 
We focused our attention mainly on NVSS radio sources from Table \ref{nvss_sources} inside the {\it Fermi} ellipse with some peculiarity from the observational point of view.
Here we comment on the most interesting cases:

i)  {\bf \object{NVSS J060716$-$061030}}. Inspection of the SPECFIND V2.0 catalog \citep{specfind} clearly indicates that this is a relatively bright radio emitter
with a well-characterized non-thermal emission. Its radio emission can be represented
by $S_{\nu} = (179\pm 1)~{\rm mJy} \left[\nu/{\rm GHz}\right]^{-0.659 \pm 0.003}$ reminiscent of a blazar spectrum. 
Unfortunately, little observational data is available
for this object  that can iestablish its nature and possible association with the {\it Fermi} LAT source. 
No optical, infrared nor X-ray counterpart is evident.

ii) {\bf \object{NVSS J060728$-$061359}}. Although no other radio detections are available, this radio source has a clear optical counterpart
candidate within the NVSS position error of about $\pm4^{\prime\prime}$. It has a clear stellar appearance with
$V=16.6$, $V-R=0.9$, $R-I=0.0$, and $3.2\pm 0.4$\% polarization \citep{j1994}. Unfortunately,
no spectroscopic or X-ray extinction information is available for dereddening and classification purposes.

iii) {\bf \object{NVSS J060731$-$062356}}. This extended radio source is in the direction of a bright, early-type star (\object{BD$-$06 1415}) surrounded by a reflection nebula \citep{rn}.
 With a visual $V=10.4$ magnitude, the spectral type is estimated as B1 \citep{obcat}. No radio detections at other frequencies have
been reported that would enable a spectral index estimate.  

iv)  {\bf \object{NVSS J060757$-$061347}}. This is a weak NVSS radio source with a large position error. It is consistent, within astrometric uncertainty, with
the point-like infrared source \object{2MASS J0607582$-$061353} also detected by {\it Chandra}  \citep{Chandra_pasj}
who satisfactorily fitted its X-ray spectrum with a thin thermal plasma model compatible with a YSO emission.
Together with the Mon R2 YSOs, this is the
only radio source with an X-ray association, although other NVSS/Chandra coincidences are not strictly ruled out because of ACIS lack of coverage.

\section{Conclusions}

   \begin{enumerate}
     
     \item We have presented an extended analysis of {\it Fermi} LAT data for the gamma-ray source 2FGL~J0607.5$-$0618c
     in the Mon R2 massive star forming region. The reality of this object as a gamma-ray emitter is confirmed at about the 12$\sigma$ level.
     With the data available so far, no clear indication that it is a variable source has been found.
     
     \item A multi-wavelength study at lower energies has been carried out to better constrain the physical parameters that will be used 
     in a later theoretical interpretation. In particular, we report a high-fidelity 6-cm map of the thermal radio emission from the Mon R2 central cluster
     obtained using a cross-calibration technique with a strong OH maser in the field.
     
     \item Different physical scenarios have been considered for 2FGL~J0607.5$-$0618c.
     If this source is simply a passive cosmic ray target, we estimate that an  enhancement 
     of the local cosmic ray energy by a factor as high as $\sim 7$ is needed. This would also
     imply the need of local cosmic ray accelerators. This strong requirement could be alleviated 
     if 2FGL~J0607.5$-$0618c is a multiple, extended source. In this case, the emission could be
     dominated by the IRS3 massive YSO, together with likely contributions from other relatively young
     objects such as T Tauri and runaway stars commonly found in star forming regions. Finally, the possibility
     of a background/foreground counterpart unrelated to Mon R2 has been considered, but no convincing
     candidate is currently available.

        \end{enumerate}

\begin{acknowledgements}
The authors acknowledge support of different aspects of this work
 by grants AYA2010-21782-C03-01 and AYA2010-21782-C03-03 from the Spanish Government, 
Consejer\'{\i}a de Econom\'{\i}a, Innovaci\'on y Ciencia
of Junta de Andaluc\'{\i}a as research group FQM-322 and excellence fund FQM-5418, as well as FEDER funds. GER is supported by
PIP 0078 (CONICET) and PICT 2007-00848, Pr\'estamo BID (ANPCyT), as well as by the Spanish Ministerio de Inovaci\'on
y Tecnolog\'{\i}a under grant AYA 2010-21782-C03-01.
J.A.C was supported by grant PICT 2008-0627 from ANPCyT and PIP 2010-
0078 (CONICET).
P.M-A acknowledges financial support from Universitat de Barcelona through an APIF fellowship.
J.M.P. also acknowledges financial support from ICREA Academia.
The NRAO is a facility of the NSF operated under cooperative agreement by Associated Universities, Inc.
Based on observations made with ESO Telescopes at the La Silla Paranal Observatory under programme ID 66.C-0310.
\end{acknowledgements}

\end{document}